\documentclass[acmtog, authorversion]{acmart}

\usepackage{booktabs} 
\usepackage{dsfont}

\usepackage{soul}
\usepackage{xcolor}
\usepackage{amsmath}

\def\figurePath{images/}
\def\myfigure#1#2{\begin{figure}[t]\centering\includegraphics*[width = \linewidth]{\figurePath#1}\caption{#2}\label{fig:#1}\vspace{-5pt}\end{figure}}

\def\mycfigure#1#2{\begin{figure*}[t]\centering\includegraphics*[clip, width = \linewidth]{\figurePath#1}\caption{#2}\label{fig:#1}\vspace{-6pt}\end{figure*}}

\newcommand{\refTbl}[1]{Table~\ref{tbl:#1}}

\definecolor{unsurecolor}{rgb}{1,.85,.7}
\definecolor{changedcolor}{rgb}{.85,1,.7}

\soulregister\ref7
\soulregister\cite7
\soulregister\citeetal7
\soulregister\etal0
\soulregister\eg0
\soulregister\scene1
\soulregister\refTbl1

\DeclareGraphicsExtensions{.pdf,.ai,.psd,.jpg,.png}
\DeclareMathOperator*{\argmin}{argmin}

\setcitestyle{square}

\usepackage[ruled]{algorithm2e} 

\SetAlFnt{\small}
\SetAlCapFnt{\small}
\SetAlCapNameFnt{\small}
\SetAlCapHSkip{0pt}
\IncMargin{-\parindent}

\acmJournal{TOG}
\acmYear{2018}
\acmVolume{37}
\acmNumber{6}
\acmMonth{11}
\acmArticle{192}

\setcopyright{acmcopyright}

\acmDOI{10.1145/3272127.3275066}

\ccsdesc[500]{Computing methodologies~Reflectance modeling}
\ccsdesc[500]{Computing methodologies~Texturing}
\ccsdesc[300]{Computing methodologies~Image processing}
\ccsdesc[300]{Computing methodologies~Shape analysis}
\ccsdesc[100]{Computing methodologies~Neural networks}

\begin{document}
\title{PhotoShape: Photorealistic Materials for Large-Scale Shape Collections
} 

\author{Keunhong Park}
\orcid{}
\affiliation{%
  \institution{University of Washington}
  \streetaddress{185 W Stevens Way NE}
  \city{Seattle}
  \state{WA}
  \postcode{98195}
  \country{USA}}
  
\author{Konstantinos Rematas}
\orcid{}
\affiliation{%
  \institution{University of Washington}
  \streetaddress{185 W Stevens Way NE}
  \city{Seattle}
  \state{WA}
  \postcode{98195}
  \country{USA}}
  
\author{Ali Farhadi}
\orcid{}
\affiliation{%
  \institution{University of Washington}
  \streetaddress{185 W Stevens Way NE}
  \city{Seattle}
  \state{WA}
  \postcode{98195}
  \country{USA}}
\affiliation{%
  \institution{Allen Institute for AI}
  \streetaddress{2157 N Northlake Way Suite 110}
  \city{Seattle}
  \state{WA}
  \postcode{98103}
  \country{USA}}

\author{Steven M. Seitz}
\orcid{}
\affiliation{%
  \institution{University of Washington}
  \streetaddress{185 W Stevens Way NE}
  \city{Seattle}
  \state{WA}
  \postcode{98195}
  \country{USA}}

\renewcommand\shortauthors{Park, K. et al}

\begin{abstract}
Existing online 3D shape repositories contain thousands of 3D models but lack photorealistic appearance. We present an approach to automatically assign high-quality, realistic appearance models to large scale 3D shape collections.  The key idea is to jointly leverage three types of online data -- shape collections, material collections, and photo collections, using the photos as reference to guide assignment of materials to shapes.  By generating a large number of synthetic renderings, we train a convolutional neural network to classify materials in real photos, and employ 3D-2D alignment techniques to transfer materials to different parts of each shape model.  Our system produces photorealistic, relightable, 3D shapes (PhotoShapes).
\end{abstract}

\keywords{image-based modeling, shape analysis, appearance transfer, texture}

\begin{teaserfigure}
  \includegraphics[width=\textwidth]{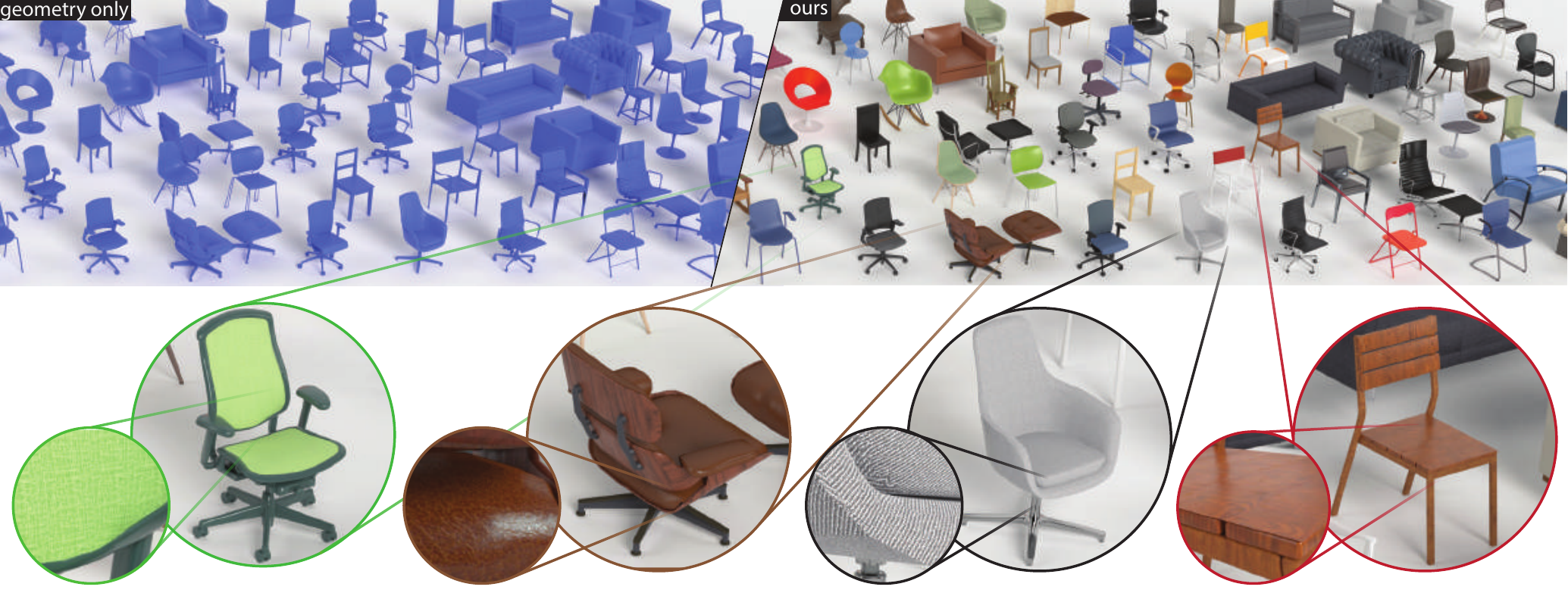}
  \caption{Fully automatic texturing of 3D shapes with rich SV-BRDF reflectance models.}
  \label{fig:teaser}
\end{teaserfigure}

\maketitle

\footnotetext{A high-resolution version of this paper along with code and data will be made available at \hyperlink{https://keunhong.com/publications/photoshape/}{https://keunhong.com/publications/photoshape/}}

\section{Introduction}

3D shapes with photorealistic materials are of great importance for problems ranging from augmented reality to game design to e-commerce.  Creating realistic 3D content is quite difficult however, and the vast majority of existing models are manually authored.

Even more difficult than producing the geometry, which an artist can author using CAD tools, is creating realistic {\em relightable textures}, as spatially varying reflectance models (SVBRDFs) are 6-dimensional.  And while computer vision-based 3D reconstruction research has advanced considerably over the last few years, existing commercial tools produce raw geometry but lack relightable textures and hierarchical part segmentation.  As a result, relatively few photorealistic relightable 3D shapes (PhotoShapes) exist, and even fewer are freely available online.

Our goal is to produce thousands of freely available PhotoShapes.  To this end, we observe that the problem of creating PhotoShapes
can be factored into three more tractable subproblems:
\begin{enumerate}
\item we need thousands of good shape models.
\item we need databases of high-quality spatially-varying material models.
\item we need an assignment of materials to shape parts.
\end{enumerate} 
The first problem (shape models) is addressed in part by the existence of large model databases like ShapeNet~\cite{shapenet2015} which captures thousands of chairs (and many other categories) spanning myriad shapes and styles.  The second problem (appearance models) is addressed in part by existing BRDF/SVBRDF databases and other online material libraries, although many of these are not free (e.g., \cite{adobe-stock}), and these collections are not as extensive as we would like; we therefore contributed a number of high quality SVBRDFs to round out the collection.  The third problem -- assignment of shape to materials -- is the focus of our paper.

Given a 3D shape of a chair with a set of parts (e.g., legs, seat, back) and a set of materials (e.g., different types of wood, plastic, leather, fabric, metal) how should we decide which materials to apply to each part?  Whereas an artist would choose this assignment manually, we propose to automate this process by leveraging {\em photos of chairs} on the Internet.  I.e., the goal is to use photos of {\em real} objects as reference to automate the assignment of materials to 3D shapes.  

Note that this problem is different than transferring the {\em texture} of a reference photo to a 3D shape, e.g., \cite{Wang2016}, as the latter requires generating missing texture for the $>50\%$ of object surfaces not visible in the photo, and does not enable specular {\em relighting} (required for many applications).  In addition to solving both of these problems, our approach produces ``super-resolution'' textures (based on the high-res appearance database) where you can zoom in far beyond the resolution afforded by the reference photo, to see fine wood-grain or stitch patterns up close.  The caveat is that these textures are ``hallucinated'' i.e., they are best matches from the database of textures rather than exact reproductions of the reference object.  This means for example, that while the overall look and appearance of the material is often matched well (e.g., ``oak'', ``black leather'') and the level of realism is high, the particular knot placement of the wood grain or the texture of the leather may differ significantly from the object in the photo. 

Conceptually, we could solve this problem by comparing every reference photo with every 3D shape, textured with every material in the database, and rendered to every viewpoint and with different illuminations.  The {\em good} matches (for each shape part) would yield our desired assignment of materials to shapes.
Aside from the obvious scale and combinatorial complexity problems with such an approach, a key challenge is how to robustly 
{\em compare} two images where features differ in both shape (e.g., arm height in an office chair) and appearance (e.g., different wood grain).  We leverage deep networks, 
trained on thousands of synthetic renderings of BRDFs applied to 3D shapes to produce robust classifiers that map patches of reference photos to material database instances.
We then use fine-scale image alignment techniques and spatial aggregation (CRFs) to assign materials to parts of the shape.

Our fully automatic system is able to produce 2,000 PhotoShapes that very accurately reflect their exemplars, and 9,000 PhotoShapes which deviate slightly but are good representations of their exemplars. In total, our system produces 11,000 photorealistic, relightable 3D shapes.

\section{Related Work}

\subsection*{Material Capture and Representation}
A widely used representation for opaque materials is the Bidirectional Reflectance Distribution Function (BRDF) and its spatially varying form (SVBRDF). Estimating BRDF parameters from images is a well studied problem, either in a lab setup~\cite{Matusik2003} or directly from images. The work of~\cite{Lombardi2015, Oxholm2016} optimize for the BRDF parameters given the shape or the illumination respectively. Chadraker etal~\cite{Chandraker2014} study how motion cues can assist the estimation. More recent approaches use deep learning to estimate the material parameters~\cite{GeorgoulisPAMI2017} from reflectance maps~\cite{RematasCVPR2016}, from the image directly~\cite{wang2017jointwild, Liu2017} or from a RGB-D sequence~\cite{kim2017lightweight}. For SVBRDFs, \cite{Aittala2013} introduces a system for easy capture of SVBRDF parameters. \cite{Dong2014} infers the diffuse and specular albedo from a moving object. \cite{Zhou2016} proposes a method for capturing SVBRDFs from a small number of views. The work of~\cite{Aittala2015} uses two photos of the same texture on a flat surface, one taken with a flash and one without. \cite{LiSiggraph16} is able to estimate the diffuse albedo, the specularity, and the illumination of a flat texture using a self-augmented neural network.

\paragraph{Diffuse Textures}
Diffuse textures maps is another technique for material representation. These textures 
can be directly applied to their corresponding 3D models~\cite{Debevec1996}. The work of \cite{OM3D2014} uses projective texturing followed by texture synthesis in 3D, after the manual alignment of a 3D object to a 2D image. \cite{Huang2018} proposes an approach for detailed geometry and reflectance extraction from single photos using rough 3D proxies. \cite{TexSynthDiamanti2015} proposes an exemplar based synthesis approach that incorporates 3D cues such as normals and light direction. 
In a different manner, \cite{Kopf2007} synthesizes solid textures that were optimized to match the statistics of a 2D image. 
Apart from their 3D applications, textures have been studied extensively in the image domain. Non-Parametric texture synthesis methods~\cite{Efros1999, Efros2001} are able to generate plausible textures from small patches. \cite{Hertzmann2001} introduced a framework that transfers the texture effects that relate two images to new one. Recently, deep learning approaches have boosted the quality of the produced textures, synthesizing either from exemplars~\cite{Gatys2015b, Sendik2017} or semantic labels~\cite{pix2pix2016, ChenK17}.

\subsection*{Material Recognition} In many computer vision applications it is important to recognize the materials that appear in an image. The method of~\cite{SharanIJCV13} uses features based on human perception for material classification. \cite{Schwartz2015AutomaticallyDL} proposes a method for discovering attributes suitable for material classification, while~\cite{zhang2015reflectance} identifies materials based on their reflectance. Similarly,~\cite{georgoulis2017material} classifies materials based on reflectance maps.
\cite{bell15minc} introduces a large dataset and a deep learning framework for material recognition in the wild (e.g. the Flickr Material Database~\cite{SharanJoV14} contains 100 images per class, with the images not beeing representative of everyday scenes). Another dataset for surface material recognition is presented in~\cite{Xue2017DifferentialAI}, together with a classification network based on differentiable angular imaging. Moreover,~\cite{wang2016dataset} introduces a light-field dataset for material recognition. \cite{cimpoi14describing, cimpoi15deep} introduces a texture dataset and a deep learning approach for texture recognition and segmentation, but the texture classes are based on high level attributes.

\subsection*{Image and Shape Dataset Analysis} Wang et al.~\cite{Wang2013} proposes a method to transfer image segmentation labels to 3D models by aligning the projections of the 3D shapes with annotated images. \cite{Huang2015} performs single-view reconstruction by jointly analyzing image and shape collections. Starting from a 3D dataset with material annotations,~\cite{Jain2012} introduces a method for material suggestions based on material relation of object parts. Similarly,~\cite{Chen2015} proposes a framework for automatic assignment of materials and textures for indoor scenes based on a set of rules learnt from an annotated database. The work of~\cite{izadinia2017im2cad} proposes a framework to infer the geometry of a room from a single image, but the appearance of the 3D models are estimated as the mean diffuse color of the image pixels. \cite{RematasPAMI2017} aligns 3D models with images to estimate reflectance maps (orientation dependent appearance), and then uses them for shading.
Closer to our work is the method of~\cite{Wang2016}, which transfers textures from images to aligned shapes. However, the transferred textures consist of only a diffuse albedo and need to contain strong patterns. Our work attempts to alleviate this limitation by using rich multi-component representations that capture a large variety of materials.


\section{Dataset}
\label{sec:dataset}

In this section we describe the three types of datasets that we used in our paper: shape, photo, and texture collections. For this work we have focused on chairs (including a variety of sofas, office chairs, stools etc.). Chairs have a diverse set of appearances and material combinations that make them appealing for our experiments.

\subsection{3D Shape Collections}
The 3D models to be textured come from two free online CAD sources: ShapeNet~\cite{shapenet2015} and  Herman Miller~\cite{hermanmiller}. In particular, we used 5,740 3D models from ShapeNet and 90 models from Herman Miller. ShapeNet is a large database of 3D models, containing thousands of 3D models across different categories. The furniture classes that are investigated in this paper are among the most populous, providing a good sampling of the ``furniture'' geometry. The Herman Miller database contains a small number of 3D models, but with higher quality meshes.

Our 3D models are in OBJ format and are segmented into parts. These parts do not always correspond to semantic groups like ``chair leg'' but are a byproduct of the 3D shape design process. Some shapes do include material information either as simple Blinn-Phong parameters or as textures, but such materials are usually low quality and inadequate for photorealistic rendering.

\paragraph{Filtering} The initial 3D shapes vary in terms of geometric detail and quality, etc. To ensure that the shape collection contains sufficient geometric quality we pre-process the database with the following steps. Firstly, we manually remove the 3D models that they do not belong to the aforementioned categories. Moreover, we remove unrealistic shapes and shapes with poor quality. Next, we delete vertex doubles and we enable smooth shading. Finally, all the models are resized to fit in a unit bounding cube.




\paragraph{UV Map Generation} Most models lack UV mappings which are required for texturing. To estimate the UV maps we use Blender's ``Smart UV projection'' algorithm \cite{vallet:wysiwig:09} for each material segment.

\subsection{Exemplars: Photographic References for PhotoShapes}
We pair 3D shapes with photographic references which we call \emph{exemplars} to guide the appearance of PhotoShapes. Our collection of exemplar consists of of 40,927 product photos that were collected from a) the Herman Miller website, and b) image search engines (Google and Bing), similar to \cite{Huang2015}. Specifically, we used 1820 images from Herman Miller and 39107 from the search engines. Product images are suitable for our task because objects of interest are uncluttered and easily segmented from the background (which is usually white). To ensure that images are unique, we remove duplicates by computing dense HOG features \cite{Felzenszwalb2010} on each image and removing images with an L2 distance lower than 0.1. Moreover, a foreground mask is computed for each image using a simple pixel value threshold. The object is then tightly cropped with a square bounding box and resized to 1000x1000. 





\subsection{Material Collection}
\myfigure{materials}{Examples of materials from each class. Rendered with Blender.}

Our goal is to provide realistic, physically based textures to a 3D model collection. Textures are represented as SVBRDFs with spatially varying diffuse, specular and roughness maps. They also contain geometric information via normal and height maps. This representation enables realistic reproduction of materials and seamless incorporation into any modern physically based renderer.

\mycfigure{pipeline_overview_v5}{An overview of our system. 
(a) The input to our system is a collection of images, shapes, and materials. (\autoref{sec:dataset}), 
(b) we take the shape and image collections and correlate them in an alignment step (\autoref{sec:alignment}), 
(c) we take each shape-image pair along with the finely aligned segmentation mask and predict the material of each part (\autoref{sec:material-classification}), (d) the output of our system is a large collection of richly textured 3D shapes (novel viewpoints are shown in the figure).}

We collected SVBRDFs by scanning real surfaces using \cite{Aittala2015} and by collecting synthetic textures from online repositories. Specifically, we manually scanned 33 materials in addition to the 34 materials provided in \cite{Aittala2015}. We downloaded 68 materials from \cite{poliigon}, 57 materials from VRay Materials~\cite{vray-materials} and 238 materials from \cite{adobe-stock}. We also manually created 15 metals and plastics.

Materials scanned using \cite{Aittala2015} were converted from their model (similar to BRDF Model A from \cite{brady2014genbrdf}) to an anisotropic Beckmann model in order to render using Blender. All the other BRDFs were rendered using their designed BRDFs. Poliigon and V-Ray Materials are rendered using the GGX \cite{Walter2007} model and Adobe Stock materials are rendered using the Disney Principled BRDF \cite{burley2012physically}.  In total, our database consist of 48 leathers, 154 fabrics, 105 woods, 86 metals, and 60 plastics. Examples of the materials are shown in \autoref{fig:materials}.

\paragraph{Normalizing Scale}
Materials are scanned or created with an arbitrary and unknown scale. In order to use materials consistently, we manually assign a scale value $s_i$ for each material $m_i \in \mathcal{M}$. This value is used as a scaling factor for the UV mappings which are scaled by a factor of $\log s_i$ during rendering.

\paragraph{Environment Maps} We also have a small set of 30 HDR environment maps from \cite{Debevec1996}, \cite{deviant-art-envmap}, and \cite{adobe-stock}. We select environment maps that simulate studio-like lighting conditions as use them for all of our renderings.

\section{Shape-Image Alignment}
\label{sec:alignment}

\myfigure{image_shape_retrieval_v2}{The top-5 shape and pose retrievals given an image (outlined).}
\myfigure{shape_image_retrieval_v2}{The top-5 image retrievals (outlined) given a shape. Computed using a reverse-index.}

Given a collection of uncorrelated 3D shapes and images of the same category, we wish to synthesize realistic textured versions of each 3D shape. To achieve this, we propose a system to extract appearance information from the images and transfer that information onto the shape collection. We pose the problem as a classification problem in which the goal is to assign a material model from our texture dataset to each 3D shape part.  Our system is comprised of two parts:

\begin{enumerate}
    \item \textit{Coarse step}: assigns to each shape a list of exemplars and associated camera poses
    \item \textit{Fine step}: creates a pixel-wise alignment between shapes and exemplars
\end{enumerate}

We use the following terminology throughout the paper: a \textit{shape} is a  3D model obtained from an online shape collection. Each shape is by construction divided into \textit{object parts} defining structural divisions (seat, arm, leg, etc.) and \textit{material parts} ($\mathcal{P}$) defining which objects should share the same material.

In order to texture a 3D shape, we refer to a set of associated image exemplars and use them as a proxy for reasoning about plausible textures for the shape. For this to be possible we must first compute an association between the collection of 3D shapes and exemplar images. We call this task \textit{alignment} and break it down into two steps: 1) a \textit{coarse} step, and 2) a \textit{fine} step.

\mycfigure{alignment_refinement_figure_v2_horizontal}{We refine a coarsely aligned mesh segment ID map to better align with the image. The shape segmentation map projected onto the image (a) after coarse alignment; (b) after applying the flow field computed using SIFT Flow; (c) after applying our dense CRF.}

\subsection{Coarse Step: Shape to exemplar matching.}
\label{sec:coarse-step}

We seek a list of exemplars for each 3D shape, as well as the camera pose for every such shape-exemplar pair. We pose coarse alignment as an image retrieval problem, solving shape retrieval and pose estimation simultaneously. Similar approaches are taken in \cite{Huang2015} and \cite{Wang2016}. For efficiency, we solve this by creating a reverse-index from exemplars to the top $k$ shape renderings. Inverting this index gives us our desired output. 

We render each shape from various viewpoints sampled from a sphere around the object. The camera is parameterized in spherical coordinates. 50 elevation values are uniformly sampled in $\phi\in[\frac{\pi}{4},\frac{9}{16}\pi]$ and $(50\sin\phi)$ azimuth values sampled uniformly over $\theta \in [0, 2\pi)$. This results in 456 distinct viewpoints. 
\paragraph{Distance Metric} We require a distance metric in order to compare the compatibility of a rendering and an image. The alignment problem is then $M(I) = \argmin_{\phi\in\Phi, \theta\in\Theta, s\in|\mathcal{S}|} \lVert F(R(s;\phi,\theta)) - F(I) \rVert$ where $M$ is a function that returns the top match, $I$ is the image query, $R$ is the renderer, $\mathcal{S}$ is the set of all shapes, $\Phi$ is the set of all elevations, $\Theta$ is the set of all azimuths, and $F$ is a feature descriptor.

We use the HOG descriptor from \cite{Felzenszwalb2010} as $F$. This feature descriptor has the benefit of low dimensionality making computation and comparisons extremely efficient. We compute descriptors of size 100x100 with a bin size of 8 yielding 1352 dimensional features. The input image is blurred with a Gaussian filter with $\sigma=0.1$ in order to reduce texture effects.

During comparison, the rendering $R(s;\phi, \theta)$ and the image $I$ are both cropped with a square bounding box around their foreground masks. This allows us to perform the coarse alignment with a simple spherical coordinate camera model forgoing focal length or translation parameters.

\subsection{Fine Step: Segmentation Refinement}
\label{sec:fine-step}

3D shapes are segmented into object and material parts by their authors upon construction. We assume that any parts of the shape which have the same material label share the same apperance. One may also use the object parts as supervision for this purpose; however, we find that these tend to be over segmented and lack the symmetry found in material segments (e.g. each leg of a chair may be assigned a different material).  Given the coarse alignment we can compute a 2D material part labeling $\mathbf{p}_\text{coarse}\in\mathcal{P}^N$ for an image of size $N$ by projecting the shape parts $\mathcal{P}=\{p_1,p_2,\ldots,p_{|\mathcal{P}|}\}$ with the estimated camera pose. The effect of this is shown in \autoref{fig:alignment_refinement_figure_v2_horizontal}(a).

A naive projecting of the coarsely aligned part mask is insufficient for associating the two modalities. The mask does not perfectly align with the exemplar and thin structures such as chair legs may have zero overlap.  We use the coarse alignment as initialization and perform an additional refinement step in order to get a cleaner pixel-wise alignment of the projected part mask.

\paragraph{SIFT Flow} We use the approach from \cite{Wang2016} in which we compute a flow which warps the projected shape segment map onto the exemplar. The flow is computed by using the SIFT Flow algorithm \cite{SIFTFlow} on the silhouettes of each map. We encode the vertical and horizontal pixel coordinates into the blue and green pixels of the silhouette image (as in \autoref{fig:pipeline_overview_v5}). This prevents the SIFT Flow step from overly distorting the mask. The resulting warped mask $\mathbf{p}_\text{flow}\in\mathcal{P}^{N}$ is shown in \autoref{fig:alignment_refinement_figure_v2_horizontal}(b).




\paragraph{Dense CRF} The SIFT Flow refinement results in an overlapping but noisy segmentation. We further clean the segmentation mask by using a dense CRF \cite{DenseCRF} in the same manner as \cite{bell15minc} (Please see supplemental materials for details). The resulting part mask $\mathbf{p}_\text{crf}\in\mathcal{P}^{N}$ which we use for the rest of the system is shown in \autoref{fig:alignment_refinement_figure_v2_horizontal}(c). The aligned part mask enabled us to share information between shapes and corresponding image exemplars.

\subsection{Substance Segmentation}
\label{sec:substance-segmentation}

We first use the aligned image exemplar to infer {\em types} of materials, a.k.a. ``{\em substances}'' for each part of the aligned object. In the next section, we will convert these substances into fine-grained SVBRDFs. We segment the image and label each pixel with a substance category using \cite{bell15minc}. For our experiments we use the substances 'leather', 'fabric', 'metal', 'wood', and 'plastic'. Similar categories are mapped to a canonical category (e.g. 'carpet' to 'fabric'). All other category probabilities are set to zero and the remainder are re-scaled to sum to 1.0. This process results in a substance mask $\mathbf{q}\in\mathcal{Q}^N$ where $\mathcal{Q}=\{q_1,q_2,\ldots,q_{|\mathcal{Q}|}\}$ is the set of substance labels.  We compute a substance labeling of the shape $\mathbf{q}_\text{shape}\in\mathcal{Q}^{|\mathcal{P}|}$ by choosing the substance label that has the most overlap. Let $s_i$ be the substance label of part $i$. This process may also yield a cleaner substance segmentation of the image. See supplementary material for an example.

This process assigns a substance label to each aligned 3D shape part computed from \autoref{sec:fine-step}.


\section{Image Segment to SVBRDF}
\label{sec:material-classification}

Our objective is to assign a plausible SVBRDF to each part of a 3D shape. One approach is to extract planar patches and optimize a texture as in \cite{Wang2016} with an SVBRDF regression method such as \cite{LiSiggraph16}. However, we find that extracting patches from images yields low resolution, distorted textures which are difficult to analyze. Extracting local planar patches also loses global context which is useful for inferring glossiness.

We instead tackle this problem as a classification problem. The input is an image and a corresponding binary mask representing a single material. The output is a material label $m\in \mathcal{M}$ chosen from our collection of SVBRDFs.   Ideally we would have a collection of real images with ground truth SVBRDF labelings. In practice, it is difficult to define such a task. We also found human judgment of reflectance (as in \cite{bell13opensurfaces}) to be noisy and low quality. We therefore generate synthetic data where we know ground truth.

\subsection{Synthesizing Training Data}
\label{sec:synthetic-rendering}
Synthetic data has shown to be effective for training or augmenting models that generalize to real world applications \cite{Su_2015_ICCV,richter2016playing}. We therefore sidestep the difficulty of collecting ground truth by creating our own.  Given our 3D shape and material databases, we can create a large amount of training data by applying different materials to shapes and rendering to a range of camera viewpoints under different illuminations. 

\paragraph{Camera Pose Prior} We find that there is a strong bias in camera poses in real images (e.g., chairs are rarely photographed from below). We thus uniformly sample from the distribution of camera poses obtained in the \hyperref[sec:coarse-alignment]{coarse alignment step}. 

\paragraph{Substance Prior} Substances do not occur randomly in objects. Legs of chairs are usually not made of leather and a sofa is usually not upholstered with metal. We leverage the shape substance labelings $q_{p_i}\in\mathbf{q}_{shape}$ in \autoref{sec:substance-segmentation} to enforce a substance prior. Instead of selecting a completely random material, we condition on the substance category and sample $m_{p_i}\sim U(\{m|m\in\mathcal{M}, q_m = q_{p_i}\})$.

\paragraph{Texture Scale Normalization}
Different tessellation and UV mappings can arbitrarily change the rendered scale of our textures. We normalize the UV scale for each mesh segment $S_i$ by computing a density $D_i = A_i^{\text{uv}}/A_i^{\text{world}}$ where $A_i^{\text{uv}}$ is the local UV-space surface area of the mesh and $A_i^{\text{world}}$ is the local world-space surface area. The UV coordinates for the segment is then scaled by $\frac{1}{D_i}$. This method assumes little or no distortion in the UV mapping.

\paragraph{Randomized Rendering}  To generate a single random rendering, we uniformly sample a shape-exemplar pair computed in \autoref{sec:coarse-step}. Given a pair, we 1) sample a camera pose from the distribution computed above, 2) assign a random material (SVBRDF) to each shape part conditioned on the substance label computed in \autoref{sec:substance-segmentation}, and 3) select a random environment map.

To improve classifier robustness, we further augment our data as follows: 1) Randomly jitter azimuth and elevation by $\Delta\theta \sim U([-\frac{\pi}{12},\frac{\pi}{12}])$ and $\Delta\phi\sim U([-\frac{\pi}{24},\frac{\pi}{24}])$, 
2) randomly select a field of view $\text{fov}_x\sim U([50, 60])$,
3) randomly select a camera distance $r\sim U([1.3, 1.75])$,
4) randomly scale the radiance of the environment map by $s_\text{env}\sim U([0.9, 1.2])$, 
5) randomly scale, rotate, and translate UV mappings by ($\Delta s_\text{uv}\in[-1,0.5]$, $\Delta\theta_s\in[0, 2\pi]$, and $\Delta x, \Delta y\in[0, 1]^2)$ respectively.  We use Blender to generate 156,262 renderings (examples in supplementary material).

\subsection{Material Classification}
\label{sec:material-classification}

Our synthetic dataset is generated by conditioning on \emph{substances}. We would like to be able to generate PhotoShapes with more accurate fine grained \emph{materials} (e.g., a specific 'oak', 'cherry', 'maple', instead of just 'wood'). Although material assignments were only conditioned on substance categories, the renderings of our synthetic dataset from \autoref{sec:synthetic-rendering} contain ground truth labels for which specific \emph{material} is rendered at each pixel.

We directly use the renderings and their ground truth \emph{materials} to train a classifier which predicts which materials are present in a specified image. We experimented with other methods such as color histogram matching but found that such brute-force matching approaches are not practical when comparing a large number of materials. Our classifier is a feed-forward neural network which is efficient even for a large number of materials.

The input to our classifier is an image exemplar concatenated with a binary segmentation mask. The input mask represents the portion of the image that is to be classified. For example, if the mask had non-zero- values only within the fabric upholstery (as in \autoref{fig:pipeline_overview_v5}(c)) the desired output would be a blue fabric.  The output of our classifier is an $|\mathcal{M}|$ dimensional vector $\mathbf{x}^m$ which when applied a soft-max operator becomes a discrete probability mass. We optimize this class labeling using a cross entropy loss
$$ \mathcal{L}_\text{mat}(\mathbf{x}^m, y^m_i) = - \log \left(\frac{\exp{x^m_i}}{\sum_j{\exp{x^m_j}}}\right) $$
where $y^m_i$ is the ground truth label.

\subsubsection{Multitask Learning} We find that naively training a network only on material class information generalizes poorly to real images. Intuitively this makes sense since we have given the system no information about the relative affinity between different materials e.g., it is less wrong to classify beech wood as cherry wood than it is to classify it as a black leather. We introduce an auxiliary task to our network in order to regularize our feature space and to teach the network the relative affinity between materials.

\paragraph{Substance Classification} Mis-classifying a wood material as leather is more detrimental than classifying it as a different wood. As such, we add an additional fully connected layer to our network which predicts the \textit{substance} category $q\in\mathcal{Q}$ of the input (wood, leather, plastic, etc.) The ground truth labels come directly from annotations of our material dataset. The substance is also a classification and is optimized using a cross entropy loss in a similar fashion to the material loss:
$$ \mathcal{L}_\text{sub}(\mathbf{x}^s, y^s_i) = -\log \left(\frac{\exp{x^s_i}}{\sum_j{\exp{x^s_j}}}\right) $$
where $y^s_i$ the ground truth label.

\paragraph{Combining Losses} The most straightforward way to optimize our objective is to compute a weighted sum of our loss functions:
$$ \mathcal{L}(\mathbf{x}, \mathbf{y}; \lambda) = \mathcal{L}_\text{mat}(\mathbf{x}^m, \textbf{y}^m) + \lambda\mathcal{L}_\text{sub}(\mathbf{x}^s, \textbf{y}^s) $$

for some weighting term $\lambda$. However, we found it more efficient to use the uncertainty weighted multitask loss from \cite{kendall2017multi} which defines a weighting based on learned homoscedastic (task-dependent and not data-dependent) uncertainties $\hat\sigma_m^2, \hat\sigma_s^2$. Concretely, our loss function is formulated as:
\begin{align}
    \mathcal{L}(\mathbf{x}, \mathbf{y}; \hat\sigma) &= \mathcal{L}_\text{mat}(\mathbf{x}^m, \textbf{y}^m)\hat\sigma_m^{-2} + \log\hat\sigma_m^2  \\
        &+ \mathcal{L}_\text{sub}(\mathbf{x}^s, \textbf{y}^s)\hat\sigma_s^{-2} + \log\hat\sigma_s^2
\end{align}

In practice, we optimize for the log variance $\hat{s}:=\log{\hat\sigma^2}$ as it avoids a possible divide by zero and is more numerically stable:
\begin{align}
    \mathcal{L}(\mathbf{x}, \mathbf{y}; \hat{\textbf{s}}) &=  \mathcal{L}_\text{mat}(\mathbf{x}^m, \textbf{y}^m)\exp(-\hat{s}_m) + \hat{s}_m  \\
        &+ \mathcal{L}_\text{sub}(\mathbf{x}^s, \textbf{y}^s)\exp(-\hat{s}_s) + \hat{s}_s
\end{align}

We initialize $\hat{s}_m=0.0, \hat{s}_m=-1.0$.

\subsubsection{Model Architecture and Training}

We use Resnet-34 \cite{he2016deep} with weights initialized to those pre-trained on ImageNet. We add a 4th alpha channel to the input layer of the network which represents the segment of the image we wish to classify. The associated filters of the alpha channel are initialized with a random Gaussian. The output of the network is a score for each of the $C + 1$ categories mapping to a material and a background class.

We train the network with stochastic gradient descent (SGD) with a fixed learning rate of 0.0001 until convergence (about 100 epochs). We performance standard data augmentations: random rotations, random crops, random aspect ratio changes, random horizontal flips, and random chromatic/brightness/contrast shifts.

\subsubsection{Pretraining on Real Images}
Although our network trained only on synthetic renderings generalizes fairly well to real images, we found that having natural image supervision helps the network learn a more robust model. We pre-train our network on the dataset of \cite{bell15minc} using their ground truth region polygons to generate input masks. Our data mostly has white backgrounds and we therefore interleave whole image inputs with cropped inputs with white backgrounds.

We randomly split the OpenSurfaces dataset into training and validation sets at a ratio of 9:1. We fine-tune the network initialized to weights pretrained on ImageNet~\cite{deng2009imagenet} and train using only the substance task with a learning rate of $0.0001$ until convergence. The network reaches $86.77\%$ top-1 validation precision on the our pretraining validation set.

\myfigure{classifier-substance-predictions}{The top-3 material predictions of our material classification network for the input shown on the left. (a) shows predictions when trained without substance supervision, (b) shows predictions when trained with substance supervision.}

\subsubsection{Inference}
\label{sec:inference}
Given an image we take the segmentation mask computed in \autoref{sec:fine-step} and infer a material for each segment. We find that we are able to improve material prediction performance by weighting the material prediction by the confidence of the corresponding substance category prediction.

\subsection{Generating PhotoShapes}
The final objective is to create a collection of photorealistic, relightable 3D shapes. Consider the collection of all aligned shape-exemplar pairs computed in \autoref{sec:alignment} as PhotoShapes \emph{candidates}. These candidates become PhotoShapes when each of their material parts are assigned a relightable texture. The latter step is done simply by applying \autoref{sec:material-classification} to each photo-aligned material part (using the mask from \autoref{sec:fine-step}). For our experiments, we took the top $12$ matching exemplars for each shape (ordered by HOG distance) and discarded any pairs with a HOG distance $<8$. We then applied our material classifier to the remaining candidates. This process yields 29,133 PhotoShapes. Examples are in \autoref{fig:results-figure-1col}.

\section{Experiments}

In this section, we describe implementation details, evaluations,
comparisons to prior work, limitations, and applications.

\subsection{Baselines}

\paragraph{Built-in Textures} The simplest baseline is to compare with the materials that come by default from each shape model. Most of the shapes in our dataset lack any meaningful material properties. Besides the very few with high quality textures, most shapes have either arbitrary colors or low resolution textures.
\paragraph{No Material Classification} We evaluate a version of our method without the material classification step. We assign random materials given the substance computed in \autoref{sec:substance-segmentation}. This is akin to our training data generation process.

\paragraph{Color Matching} A naive method of matching materials is to render a shape with all possible materials and compare the resulting color distribution with the exemplar image. Our experiments show that such brute-force approaches were too slow for large-scale applications (rendering thousands of object with hundreds of materials takes a long time). Such comparisons also have difficulty accounting for complicated textures and non-uniform lighting effects such as shadows and specularities. We show a comparison in \autoref{fig:median_color_comparison}.

\paragraph{Projective Texturing} Another approach for appearance modeling is projective texturing. If the alignment between the 3D model and the image is good ~\cite{Debevec1996, OM3D2014}, the image can be used as a texture for the 3D model. However, good alignment is a very difficult task, even with manual intervention. An additional approach that models the appearance of an object as ``baked'' material and illumination properties is reflectance maps \cite{RematasPAMI2017}. \autoref{fig:result-comparison} shows the results of projective texturing and reflectance map shading compared to our approach.

\myfigure{median_color_comparison}{Matching by median color fails to account for specular reflections and fetches a diffuse, lighter material. We include the color histogram of each image on the upper left with the median color (dotted lines).}

\subsection{Results}
We show a sample of our results in \autoref{fig:results-figure-1col} (please see supplementary materials for more), comparing to the built-in materials, our results with no classifier, and our full results. We also show the benefit of using high-resolution relightable textures in \autoref{fig:closeup_figure}.

We also show results for categories other than chairs in \autoref{fig:other_category_results}. Note that these results are produced without additional training. Results may be improved by adding relevant materials (e.g., rubber for tires) and further training the classifier.

\mycfigure{results_three_way}{Our method is able to produce different plausible materials (bottom) of the same Shape (left) given different exemplar images (top outlined).}

\myfigure{other_category_results}{Without any additional training, our pipeline produces plausible results for motorcycles (top left), pillows (top right), coffee tables (bottom left) and cabinets (bottom right). A green outline indicates the exemplar followed by our result to the right. The motorcycle tires are assigned metallic and leather materials as we lack a rubber material in our dataset. The knobs of the cabinet were missed by the fine alignment step due to the small size and sharp color variation of the chrome.}

\mycfigure{result-categories}{Representative examples of our categorization of results. (a) Good models are good representations of the original exemplar. (b) Acceptable models have slight differences but are overall plausible. (c) Failures, for which we identify the following modes: (i) material mis-classification, (ii) material mis-classification caused by ambiguous appearance (plastic sometimes looks like leather etc.), (iii) color mismatch, (iv) over-segmented mesh causing material discontinuity, (v) under-segmented mesh making it impossible to assign correct materials, and (vi) mis-alignment of the exemplar and shape (includes retrieving the wrong object)}

\subsection{User Study}

We conducted a user study on Amazon Mechanical Turk to evaluate the performance of our method. We showed users an image and asked them to specify whether they thought the image was a "real photograph" or "generated by a computer". We tested our method with three different baselines: built-in textures, our pipeline but without the material classifier, our full method, and real images. We performed the study on the ShapeNet and Herman Miller datasets separately. For ShapeNet results, we sampled 1000 result renderings uniformly at random. For Herman Miller results we sampled 500. Results are shown in \autoref{table:user-study-table}.

\begin{table}[H]
\centering
\caption{User Study. The ratio of users who thought our results were real photographs.}
\begin{tabular}{l|ll|ll}
\hline\hline
    & \multicolumn{2}{c|}{ShapeNet} & \multicolumn{2}{c}{Herman Miller} \\ \hline
   & Real    & Fake             &  Real & Fake  \\ \hline
Built-in      & 32\%  & 68\%    & 37\% & 63\%     \\
Ours (No Classifier) & 41\% & 59\%     & 43\% & 57\%  \\ 
Ours      & \bf{47\%}  & 53\%        & \bf{51\%} & 49\%    \\
Photographs      & 81\%  & 19\% & 83\% & 17\%   \\
\hline
\end{tabular}

\label{table:user-study-table}
\vspace*{-2mm}
\end{table}

\subsection{Shape to PhotoShapes Conversion Rate}

We also manually evaluated our results. We categorize resulting PhotoShapes from our pipeline into the following categories: good, acceptable, and failure. \textit{Good} results represent their exemplar image almost exactly. \textit{Acceptable} results have slight differences from the exemplar but are good representations. We ignore shapes with low mesh quality (very low polygon count, holes, incorrect normals) and bad exemplars (watermarks, transparent materials, etc.). Everything else are considered \textit{failures}. We identify four main failure modes: (1) material mis-classification e.g., metal instead of wood, (2) color or pattern mismatch, (3) failures due to the under- or over-segmentation of meshes, and (4) incorrect shape retrievals or mis-alignments. Representative examples of each of each category are shown in \autoref{fig:result-categories}.

We evaluate our two input shapes collections (ShapeNet and Herman Miller) separately. We generate 28,432 PhotoShapes for ShapeNet and 701 PhotoShapes for Herman Miller shapes. Since manually sorting all results is not feasible, we evaluate quality on a random sampling. We uniformly randomly sample 1,322 PhotoShapes from ShapeNet and 243 PhotoShapes from Herman Miller and categorize them into the aforementioned classes. We found most failure cases were due to mis-alignments and under-segmentations of the mesh. \autoref{table:quality-table} shows the division between each category, and \autoref{table:failure-table} shows a finer division between different failure modes. Our shapes from Herman Miller have considerably higher mesh quality, resulting in a higher success rate due to a lower number of mis-alignments.

Extrapolating from our categorization, our system was able to produce around 2,100 'good' PhotoShapes for ShapeNet and 262 'good' PhotoShapes for Herman Miller. Including 'acceptable' results, we generated 11,000 PhotoShapes for ShapeNet and 475 PhotoShapes for Herman Miller.

We also report input shape coverage (projected numbers shown in parentheses). Our annotations show that   14.93\% (856) of ShapeNet shapes and 68.42\% (69) of Herman Miller shapes had least one 'good' PhotoShapes. When we include 'acceptable' results, ShapeNet had a success rate of 64.25\% (3,687) and 91.23\% (92) for Herman Miller.

While a success rate like 64\% may not sound impressive, it is very significant in the context of the problem that we're trying to solve, i.e., generating a large dataset of high quality PhotoShapes.  I.e., {\bf we have generated over 3,500 photorealistic, relightable, 3D shapes of chairs}.  It's not critical that we texture every shape, as many of these (particularly with ShapeNet) were artist-created and may not correspond to real furniture for which photo exemplars exist.

\begin{table}[H]
\centering
\vspace*{-2mm}
\caption{Our generated PhotoShapes divided into good, acceptable, and failure categories. We also show the sum of the good and acceptable classes. See \autoref{table:failure-table} for a more detailed division of failures.}
\begin{tabular}{llll|l}
\hline\hline
              & Good    & Acceptable & Failure & Good+Acc \\ \hline
ShapeNet      & 6.08\%  & 32.76\%    & 61.15\% & 38.85\% \\
Herman Miller & 37.45\% & 30.45\%    & 32.10\% & 67.90\% \\ \hline
\end{tabular}
\label{table:quality-table}
\vspace*{-2mm}
\end{table}

\begin{table}[H]
\centering
\caption{Different modes of failure. 'align' refers to mis-alignment errors, 'subst' refers to substance mis-classification errors, 'color' refers to incorrect color or pattern predictions, 'und-seg' refers to errors due to the under-segmentation of the shape, and 'ov-seg' refers to awkward results due to models being over-segmented. See \autoref{fig:result-categories} for examples.}
\begin{tabular}{llllll}
\hline\hline
             & align & subst & color   & und-seg & ov-seg \\ \hline
ShapeNet     & 40.18\%   & 27.93\%   & 11.86\% & 17.98\%   & 2.04\%   \\
Herman Miller & 14.10\%   & 38.46\%   & 32.05\% & 14.10\%   & 1.28\%   \\ \hline
\end{tabular}
\label{table:failure-table}
\end{table}





\myfigure{closeup_figure}{Here we show some of our results and close-ups. By using SVBRDFs to model we are able to infer appearance at great detail, even if the exemplar image has low resolution.}
\myfigure{result-comparison}{Comparisons of producing a novel view of an exemplar image using projective texturing, reflectance maps, and then our method.}

\subsection{Material Classifier Evaluation}

\paragraph{Data Split} We split our synthetic rendering dataset into a training set and a validation set. Since our materials are our prediction categories, we perform the split on the set of shapes and environment maps that are used to generate our renderings. We set aside 10\% of our shapes and environment maps as validation and use the rest for training. This process yields 156,511 renderings for training and 15,872 renderings for validation.

\paragraph{Ablation Study} We evaluate our network with the following metrics: a) the top-1 material class precision, b) top-5 material class precision, c) top-1 substance class precision of the substance output layer, and d) the top-1 substance class precision of the substance predicted by aggregating the confidence of each material class by substance category.

We compare four different versions of our network: 1) trained only with material class supervision, 2) trained with material and substance class supervision, 3) pretrained on OpenSurfaces and then trained only with material class supervision, and 4) pretrained on OpenSurfaces and then trained with material and substance class supervision. The results for these metrics are shown in \autoref{table:classifier-ablation}.

Our additional substance categorization task significantly boosts the validation accuracy of our classifier. We also find that material predictions are qualitatively more robust when trained with substance supervision as shown in \autoref{fig:classifier-substance-predictions}.

\begin{table}[t]
\centering
\caption{Ablation Study. We compare different versions of our model on our synthetic validation set. We try the permutations of whether or not we pretrain on OpenSurfaces and the presence of an additional substance task. (a) mtl@1 is the top-1 validation precision,  (b) mtl@5 is the top-5 precision, (c) sub@1 is the top-1 substance precision of the substance task layer prediction, and (d) sub-mtl@1 is the top-1 precision of the substance prediction implied by the material prediction.}
\begin{tabular}{lllllll}
\hline\hline
pretrain & sub       & mtl@1            & mtl@5            & sub@1         & sub-mtl@1        \\ \hline
N        & N         & 33.87\%          & 61.95\%          & -                 & 71.38\%          \\
N        & Y         & 37.17\%          & 64.31\%          & 75.50\%          & 76.58\%          \\
Y        & N         & \textbf{37.59\%} & 64.69\%          & -                 & 72.37\%          \\
Y        & Y         & 37.34\%          & \textbf{65.45\%} & \textbf{75.51\%} & \textbf{76.60\%} \\
\hline
\end{tabular}
\label{table:classifier-ablation}
\end{table}

\subsection{Image-Based SVBRDF Retrieval}
Finding or designing a suitable material for a 3D scene may be time consuming. An application of our work is retrieving a BRDF based on an image. A user-specified region of an image may be used as input to our classifier in order to produce a ranking of BRDFs. 

Despite being provided very little natural image supervision (only pretraining on OpenSurfaces), our material classifier is able to generalize surprisingly well to general photographs that do not have white backgrounds. Examples of predictions on such images are shown in \autoref{fig:real-scene-predictions}.

\myfigure{real-scene-predictions}{Without any additional training, our network is able to make reasonable predictions on natural images. The input image and colored outline for the segmentation mask is shown on the left. The top BRDF retrieval for each segment are shown on the right.}


\myfigure{results-figure-1col}{Selected Results. (a) shows the input shape, (b) the exemplar image, (c) a rendering with the default materials that come with the shape, (d) rendering with materials sampled conditioned on substance category, (e) renderings of our final PhotoShapes. Please see supplementary materials for more results.}

\section{Discussion}

\subsection{Limitations and Future Work}

Our results are limited in part by the variety of materials available -- we cannot reconstruct the wheels of the motorcycle in \autoref{fig:other_category_results} because we do not have a rubber tire material.  Our results could be improved by expanding the material database or by exploring methods to augment the current materials. This may be done by synthetically varying color and glossiness, or by \emph{synthesis} of novel SVBRDFs.

Our work relies heavily on reliable matching of photos and shapes. Most of our failures come from mis-alignments or under-segmentations of the input shape. Adding more exemplar images and filtering low-quality shapes would yield better results. Incorporating more sophisticated alignment and segmentation methods are interesting topics for future work.

\subsection{Conclusion}

We have presented a framework that assigns high quality relightable textures to a collection of 3D models with limited material information. The textures come from a large database and the material-to-3D assignment is performed with the guidance of real images to ensure plausible material configurations, and yields thousands of high quality PhotoShapes.


\begin{acks}
We thank Samsung Scholarship, the Allen Institute for Artificial Intelligence, Intel, Google, and the National Science Foundation (IIS-1538618) for supporting this research. We thank Dustin Schwenk for his help with the user study.
\end{acks}

\bibliographystyle{ACM-Reference-Format}
\bibliography{bibliography}

\end{document}